\documentclass[twocolumn,aps,pra,superscriptaddress]{revtex4-2}
\usepackage{natbib}
\usepackage[T1]{fontenc}
\usepackage[latin9]{inputenc}
\usepackage{babel}
\usepackage{float}
\usepackage{physics}
\usepackage{amsmath}
\usepackage{amssymb}
\usepackage{amsfonts}
\usepackage{xcolor}
\usepackage{bbm}
\usepackage[caption=false,font=normalsize,labelfont=sf,textfont=sf]{subfig} % caption=true centers the caption. Looks horrible.
\usepackage{graphicx,epstopdf}
\usepackage{url}
\usepackage{textcase}
\usepackage{bm}
\usepackage[unicode=true]{hyperref}
\usepackage{listings}
\makeatletter

\newcommand{\pt}{\mathcal{PT}}

\hypersetup{colorlinks=true, linkcolor=blue, filecolor=blue, urlcolor=cyan,}
\makeatletter
\newcommand*\bigcdot{\mathpalette\bigcdot@{.5}}
\newcommand*\bigcdot@[2]{\mathbin{\vcenter{\hbox{\scalebox{#2}{$\m@th#1\bullet$}}}}}
\makeatother
\begin{document}
\title{Floquet exceptional contours in Lindblad dynamics with time-periodic drive and dissipation}
\author{John Gunderson}
\affiliation{Department of Physics, Indiana University-Purdue University Indianapolis (IUPUI), Indianapolis 46202, Indiana, USA}
\author{Jacob Muldoon} 
\affiliation{Department of Physics, Indiana University-Purdue University Indianapolis (IUPUI), Indianapolis 46202, Indiana, USA}
\author{Kater W. Murch}
\affiliation{Department of Physics, Washington University, St. Louis 63130, Missouri, USA}
\author{Yogesh N. Joglekar}
%\email{yojoglek@iupui.edu}
\affiliation{Department of Physics, Indiana University-Purdue University Indianapolis (IUPUI), Indianapolis 46202, Indiana, USA}
\date{\today}

\begin{abstract}
The dynamics of an isolated quantum system is coherent and unitary. Weak coupling to the environment leads to decoherence, which is traditionally modeled with a Lindblad equation for the system's density matrix. Starting from a pure state, such a system approaches a steady state (mixed or otherwise) in an underdamped or overdamped manner. This transition occurs at an eigenvalue degeneracy of a Lindblad superoperator, called an exceptional point (EP), where corresponding eigenvectors coalesce. Recent years have seen an explosion of interest in creating exceptional points in a truly quantum domain, driven by the enhanced sensitivity and topological features EPs have shown in their classical realizations. Here, we present Floquet analysis of a prototypical qubit whose drive or dissipator strengths are varied periodically. We consider models with a single dissipator that generate global loss (phase damping) or mode-selective loss (spontaneous emission). In all cases, we find that periodic modulations lead to EP lines at small dissipator strengths, and a rich EP structure in the parameter space. Our analytical and numerical results show that extending Lindblad Liouvillians to the Floquet domain is a new, potentially preferred route to accessing exceptional points in the transient dynamics towards the Lindblad steady state.   
\end{abstract}
\maketitle
%----------------------------------------------------------------------------------------------------------------------%

\section{Introduction}
\label{sec:intro}

More than two decades ago, Bender and coworkers discovered a broad class of non-Hermitian Hamiltonians with purely real spectra for a non-relativistic particle on an infinite line~\cite{Bender1998}. This discovery significantly broadened the class of Hamiltonians considered ``physically relevant'' from Hermitian to those that are invariant under combined operations of parity and time reversal ($\mathcal{PT}$)~\cite{Bender2001,Mostafazadeh2002,Bender2007,Mostafazadeh2010}. Two decades of subsequent progress has made it clear that such non-Hermitian, $\mathcal{PT}$-symmetric Hamiltonians can represent open, classical wave systems with balanced, spatially or temporally separated gain and loss~\cite{Joglekar2013,Pile2017,ElGanainy2018}. $\mathcal{PT}$-symmetric Hamiltonians and their non-Hermitian generalizations have degeneracies where $n$ eigenvalues and corresponding $n$ eigenvectors coalesce, giving rise to exceptional points of order $n$ (EP$n$)~\cite{Kato1995}. In presence of a dimensionless, small perturbation $\delta\ll1$, the mode-splitting that is generated at an EP$n$ scales as $\delta^{1/n}\gg \delta$, and thereby provides the enhanced sensitivity that scales with the order of the EP in its proximity. Second~\cite{Chen2017}, third~\cite{Hodaei2017}, and higher order EPs have been experimentally realized in (semi) classical systems such as optics~\cite{Miri2019}, acoustics, and photonics~\cite{Ozdemir2019,Xue2020}, where sensitivity enhancement has been demonstrated~\cite{Lai2019,Hokmabadi2019}. These works have also highlighted the role of excess noise in the proximity of the EP~\cite{Wang2020,Wiersig2020} that arises from non-orthogonal eigenvectors of non-Hermitian matrices. 

Interest in exceptional points in the quantum domain has increased rapidly in the past two years. Due to the quantum limit on noise in linear amplifiers~\cite{Haus1962,Caves1982}, $\pt$-symmetric Hamiltonians with balanced gain and loss are not possible at a quantum level~\cite{Scheel2018}. However, EPs also occur as degeneracies of mode-selective lossy Hamiltonians. The coherent, non-unitary dynamics of such Hamiltonians have been simulated with quantum photonics~\cite{Xiao2017}, ultracold atoms~\cite{Li2019}, unitary quantum simulators~\cite{Sparrow2018-Nature}, and by embedding such Hamiltonian into a larger, time-dependent, Hermitian Hamiltonian~\cite{Wu2019}. In minimal quantum systems that are governed by the Gorini-Kossakowski-Sudarshan Lindblad equation~\cite{Gorini1976,Lindblad1976}---henceforth called the Lindblad equation---such dynamics are realized by post-selecting on quantum trajectories that do not undergo quantum jumps associated with the Lindblad dissipators~\cite{Naghiloo2019,Klauck2019}. However, we note that the post-selection probability exponentially decreases with time and thereby restricts the observation window for such dynamics. 

Another direction to access EP degeneracies in minimal quantum systems is to {\it not carry out post-selection}, and instead consider all quantum trajectories including those with quantum jumps that lead to damping. At critical damping, the open quantum system is at an exceptional point. Formally, for an $N$-dimensional system (with an $N\times N$ density matrix), one investigates the degeneracies of the $N^2\times N^2$ non-Hermitian, Liouvillian matrix that is obtained by vectorizing the Lindblad equation~\cite{Minganti2019}. In this approach, one needs to carefully distinguish between exceptional points, if any, of the non-Hermitian Hamiltonian that results from post-selection and the exceptional points of the Lindblad Louivillian~\cite{Minganti2019}. It is worth pointing out that this approach to EPs in the quantum domain has its own costs. Since one is studying the transient dynamics towards a steady state, there is a finite time-window after which experimental signals fall below the detector-noise floor. In particular, at the Liovillian EP, the transient decay rate is maximum. Another salient point of distinction between the two types of EPs is that while a quantum system with sufficient control can be initialized in a chosen eigenstate of the non-Hermitian Hamiltonian, all but one density-matrix eigenstates of the Lindblad Liouvillian are traceless and unphysical~\cite{Minganti2019}. Therefore, observing quantitative signatures of a Lindblad Liovillian EP and its order (including the coalescence of its eigenmatrices) is an exceptionally challenging task.  

In this article, we investigate the EPs that emerge when the Hermitian Hamiltonian or the dissipator terms in the Lindblad density matrix equation are periodically varied. Periodically driven Hermitian systems~\cite{Hanggi1998} have attracted tremendous attention in recent years because Floquet engineering can be used to create novel states of matter or induce transitions among them~\cite{Goldman2014,Oka2019}. When a $\pt$-symmetric, non-Hermitian Hamiltonian is made time-periodic, a rich landscape of EP contours that includes lines of EPs at vanishingly small non-Hermiticity emerges~\cite{Joglekar2014,Lee2015}. For an open system described by a time-periodic, non-Hermitian Hamiltonian, the key qualitative features of the EP landscape are the same irrespective of whether the Hermitian part of the Hamiltonian or the (anti-Hermitian) gain-loss part is periodically varied. These Floquet-engineered phenomena have been experimentally observed in active~\cite{Chitsazi2017} and passive~\cite{LenMontiel2018} electrical circuits, and ultracold atoms~\cite{Li2019}. However, the effects of time-periodic drives or dissipator strengths on Lindblad dynamics, including qualitative differences between the two protocols, have not been investigated.

The plan of the paper is as follows. In Sec.~\ref{sec:model} we recall Lindblad density matrix equation for an $N$-dimensional system in superoperator form, its key properties, and the vectorization method that converts it into an equation for a $N^2$-dimensional column vector whose dynamics are governed by an $N^2\times N^2$, non-Hermitian Lindblad Liouvillian matrix $\mathcal{L}$. For time-periodic coefficients in the matrix $\mathcal{L}$, we obtain the one-period time-evolution operator $G(T)$ whose eigenvalues and right eigenvectors determine the EPs in the parameter space. In Sec.~\ref{sec:jfloquet} we present the numerically obtained phase diagrams for sinusoidal and square-wave (piecewise constant) drives with variable amplitudes. In Sec.~\ref{sec:gfloquet}, we present typical phase diagrams that are obtained by periodically varying the dissipator strength, either sinusoidally or in a square-wave form. These results are obtained for two categories, one when the post-selected non-Hermitian Hamiltonian has EPs, and another when it does not. We present analytical results for the EP contour locations and shapes for a square-wave modulation in Sec.~\ref{sec:exact}. We conclude the paper in Sec.~\ref{sec:disc} with a brief discussion. 
%----------------------------------------------------------------------------------------------------------------------%

\section{Lindblad Liouvillian model}
\label{sec:model}
When a small, $N$-dimensional quantum system is coupled to a larger system, the combined system undergoes unitary evolution under a Hermitian Hamiltonian $H=H_S+H_L+H_\mathrm{int}=H^\dagger$ where $H_S$ is the small-system Hamiltonian, $H_L$ is the large-system Hamiltonian, and $H_\mathrm{int}$ encodes the interaction between the two systems. Under weak-coupling and Markovian large-system approximations~\cite{Manzano2020}, the density matrix $\rho$ of the small system, obtained by tracing out the larger system, evolves as
\begin{align}
\label{eq:rho1} 
\partial_t\rho(t)&=-i[H_S,\rho(t)]-\sum_{k=1}^{N^2-1}\frac{\gamma_k}{2}\mathcal{D}[F_k]\rho(t),\\
\label{eq:rho2}
\mathcal{D}[F]\rho&=-\left[F^\dagger F\rho(t) +\rho(t) F^\dagger F\right]+2 F\rho(t)F^\dagger,
\end{align}
where $\gamma_k\geq 0$ are dissipator strengths, $F_k$ are (generally non-Hermitian) dissipation operators, and $\mathcal{D}$ denotes the superoperator action on the density matrix. Note that each superoperator action $\mathcal{D}[F_k]$ on the density matrix preserves its trace and positive $\gamma_k$ ensure that the density matrix eigenvalues remain non-negative. The anti-commutator in Eq.(\ref{eq:rho2}) gives rise to a (generally) mode-selective, anti-Hermitian loss term, and the $2F\rho F^\dagger$ term sandwiching the density matrix is traditionally called the quantum-jump term. Thus, in the absence of quantum jumps, or post-selecting on quantum trajectories that have not undergone a quantum jump, the density matrix undergoes a coherent evolution with a non-Hermitian Hamiltonian~\cite{Naghiloo2019} given by $H_\mathrm{nH}=H_S-i\sum_k\gamma_k F^\dagger_k F_k$. 

We vectorize the $N\times N$ density matrix $\rho$ into a column vector $|\rho^v\rangle$ by stacking its columns~\cite{Sudarshan1961}, i.e. $\rho_{mn}\rightarrow \rho^v_{m+(N-1)n}$. Using the identity $A\rho B\rightarrow (B^T\otimes A)|\rho^v\rangle$, Eq.(\ref{eq:rho1}) can be written as $\partial_t|\rho^v(t)\rangle=\mathcal{L}(t)|\rho^v(t)\rangle$ with an $N^2\times N^2$ Liouvillian matrix
\begin{align}
\mathcal{L}(t)&=-i\left[\mathbbm{1}\otimes H_S(t)-H_S^T(t)\otimes\mathbbm{1}\right]\nonumber\\
-&\sum_k\frac{\gamma_k(t)}{2}\left[\mathbbm{1}\otimes F^\dagger_kF_k+(F^\dagger_k F_k)^T\otimes\mathbbm{1}-2 F^{*}\otimes F\right],
\label{eq:rhov1}
\end{align}
where $F^*$ denotes the complex conjugate (Hermitian adjoint+transpose) of operator $F$. 

%------------------------------------------------%
%------------------------------------------------%
\begin{figure*}
\centering
\hspace{-5mm}
\includegraphics[width=0.54\textwidth]{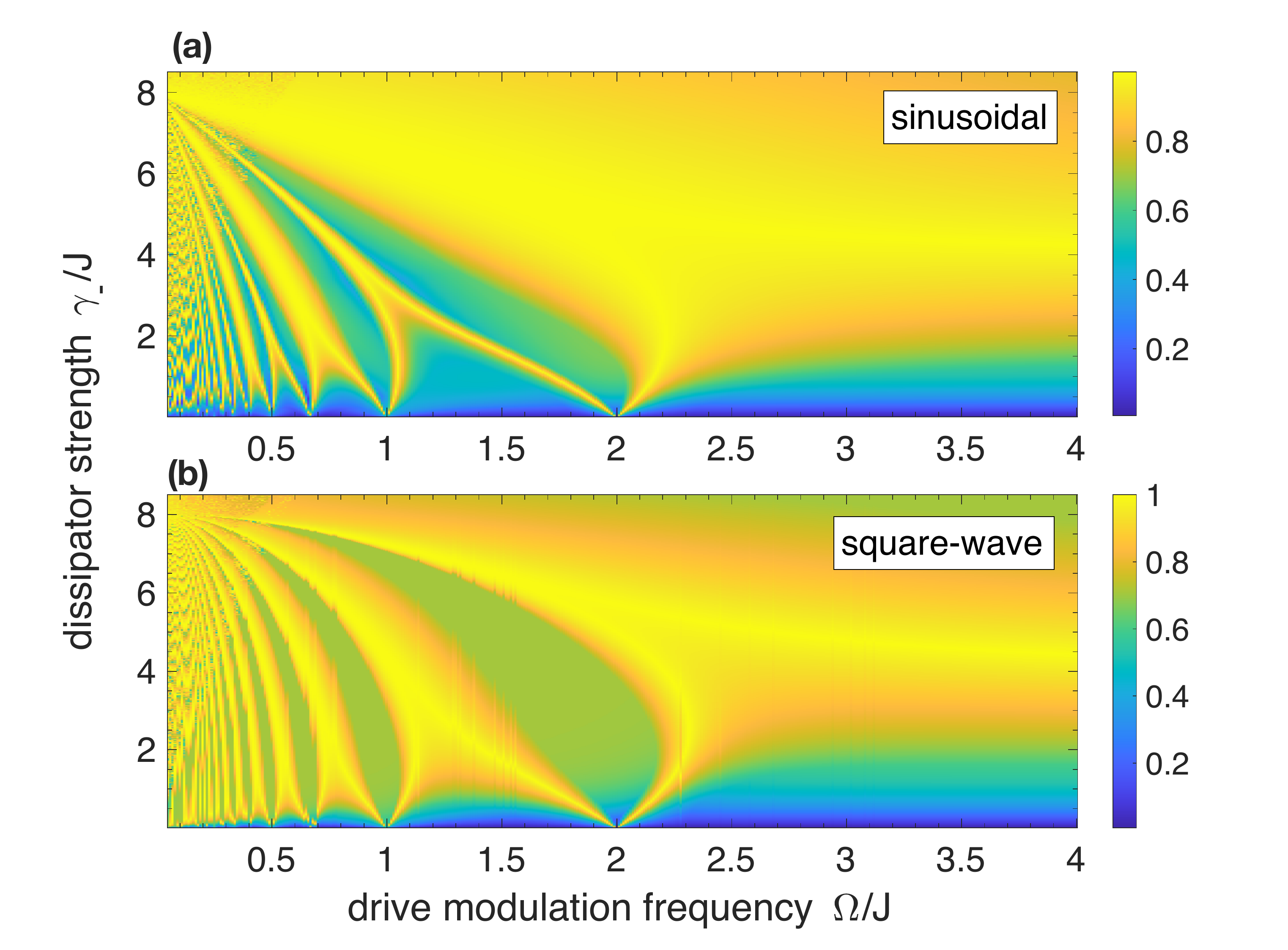}
\hspace{-7mm}
\includegraphics[width=0.51\textwidth]{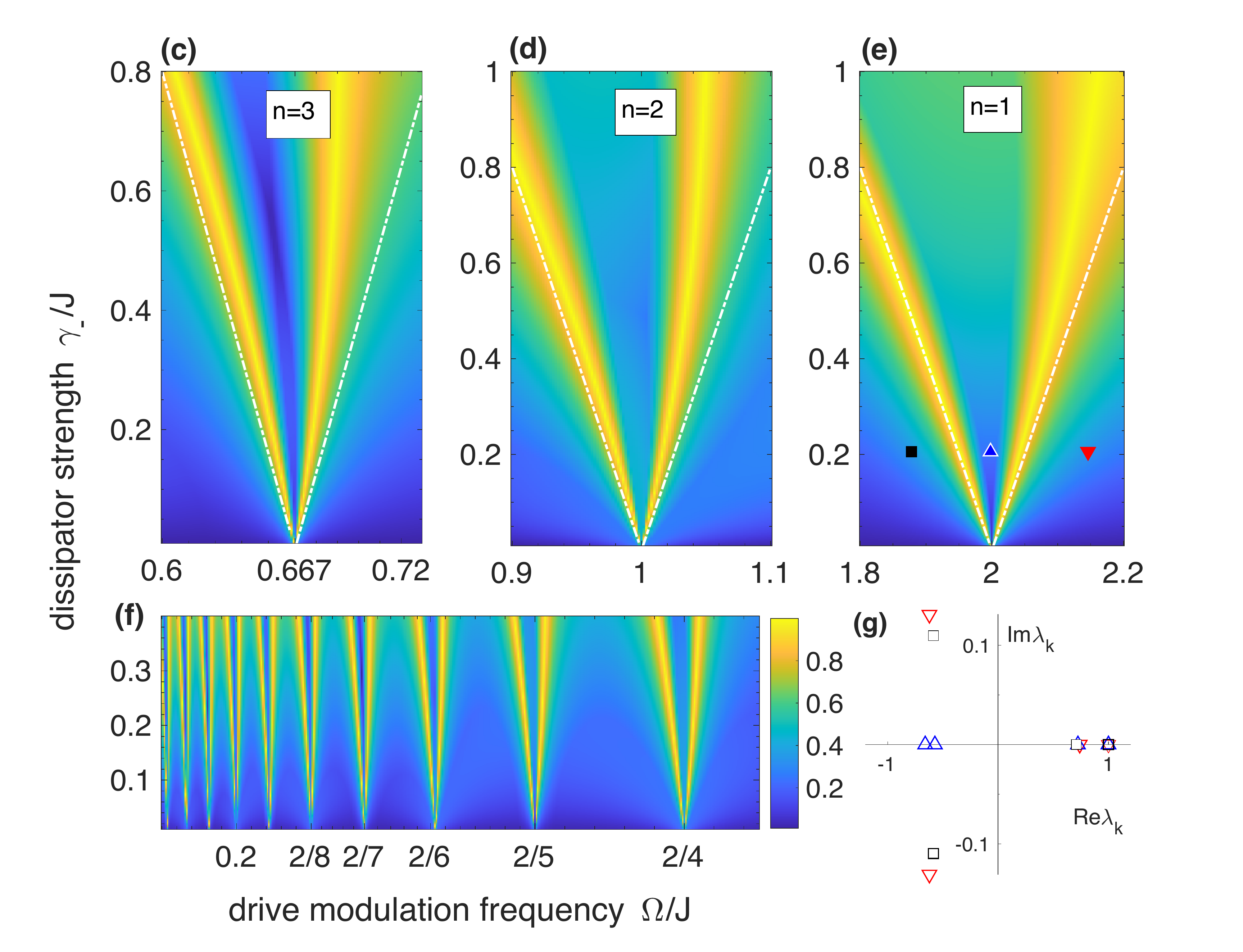}
\caption{Liouvillian EPs for a single qubit with periodic Rabi drive, Eq.(\ref{eq:jf}) and $\sigma^{-}$ dissipator with constant strength $\gamma_{-}$. (a) $\mathrm{IP}(\gamma_{-},\Omega)$ map shows the static EP at $\gamma_{-}=8J$ and a line of EPs in the high-frequency limit at $\gamma_{-}=4J$. EPs also occur along the modulation-frequency axis at all subharmonics $\Omega_n=2J/n$ for $n\geq 1$. (b) $\mathrm{IP}(\gamma_{-},\Omega)$ map for square-wave modulation gives qualitatively same results including lines of EPs that emerge from $\Omega_n$ at small dissipator strength. (c)-(e) Detail of the $\mathrm{IP}$ map in the vicinity of $\Omega_3=2J/3$ (c), $\Omega_2=J$ (d), and $\Omega_1=2J$ (e) show lines of EPs emerging from $\Omega_n$ with slopes $\pm4n$ (white dot-dashed lines). (f) Detail of lower frequencies shows ten subharmonic resonances at $\Omega_n$ for $3\leq n\leq 13$. (g) Evolution of $\lambda_k$ at $\gamma_{-}/J=0.4$ and three $\Omega$ values marked by triangles in (e). At $\Omega=2J$, all $\lambda_k$ lie on the real axis and indicate overdamped behavior. At $\Omega/J=1.95$ and $\Omega/J=2.17$, two eigenvalues are in the complex plane, indicating underdamped approach to the equilibrium state.}  
\label{fig:jfloquet1}
%\vspace{-3mm}
\end{figure*}
%-------------------------------------------------%
%-------------------------------------------------%

When $\mathcal{L}$ is static, its eigenvalues, eigenmatrices, and exceptional points have been extensively studied in Ref.~\cite{Minganti2019}. When the Liouvillian is periodic in time, $\mathcal{L}(t+T)=\mathcal{L}(t)$ for some period $T$, density matrix at long times is determined by the time-evolution operator for one period, 
\begin{equation}
\label{eq:gt}
G(T)=\mathbbm{T} e^{\int_0^T\mathcal{L}(t')dt'}\equiv e^{T\mathcal{L}_F}
\end{equation}
and the micromotion that can occur within one period. Note that the ``Floquet Liouvillian'' $\mathcal{L}_F$ is defined by Eq.(\ref{eq:gt}); whether it is a conditionally completely positive map (that ensures Markovianity) or not is a subtle question~\cite{Wolf2008,Dai2016,Schnell2020}. Instead, we investigate the eigenvalues $\lambda_k$ and eigenmatrices $|\rho^v_k\rangle$ of the unambiguously defined time-evolution operator $G(T)$. Due to the Hermiticity constraint on the density matrix, the eigenvalues $\lambda_k$ are either purely real or occur in complex conjugate pairs.  Due to the transient nature of the dynamics they represent, it follows that $|\lambda_k|\leq 1$ where the equality is only satisfied for steady-state solutions. We will use $|\rho^v_0\rangle$ to denote the physical, steady-state eigenmatrix with unit trace, and $|\rho^v_k\rangle$ ($k=1,\ldots,N^2-1$) to denote the remaining trace-zero eigenmatrices. We obtain the exceptional points of $G(T)$ by tracking the inner-product ($\mathrm{IP}$) between different trace-zero eigenmatrices, 
\begin{equation}
\label{eq:ip}
\mathrm{IP}=\max_{m>n\geq 1} |\langle \rho^v_m|\rho^v_n\rangle|\geq 0.
\end{equation}
Note that $\mathrm{IP}=1$ marks the coalescence of two (or more) eigenmatrices. When an EP of order 3 or higher is expected, such as in the case with a detuned, dissipative qubit~\cite{AmShallem2015}, we can identify it by constructing a refined metric that averages the inner product magnitudes. 
%----------------------------------------------------------------------------------------------------------------------%

\section{Liouvillians with a periodic drive}
\label{sec:jfloquet}

Consider a single qubit ($N=2$) with a Rabi drive whose amplitude is varied periodically, and a single dissipator~\cite{Minganti2019}. The complete set of $N^2-1=3$ orthogonal dissipators in this case is given by $F_k=\{\sigma_{+},\sigma_{-},\sigma_z\}$ and they represents spontaneous emission, absorption, and phase-noise processes respectively. We consider the Hermitian Hamiltonian for a resonantly driven qubit, $H_S=-J(t)\sigma_x$, with two different periodic functions for $J(t)$, 
\begin{equation}
\label{eq:jf}
J(t)=\left\{\begin{array}{c}
(J/2)[(2-\delta)+\delta\cos(\Omega t)],\\
J\,\mathrm{Square}(\Omega t),
\end{array}\right.
\end{equation}
where $\Omega=2\pi/T$ and $\mathrm{Square}(\Omega t)$ is a square-wave switching between 1 and 0. The pure-tone cosine modulation gives rise to a tridiagonal Floquet Hamiltonian in the frequency space, and is thus useful for frequency-space calculations. The piecewise-constant square modulation contains all frequencies, but is useful for analytical, time-domain calculations. First, we consider $F=\sigma_{-}$ with strength $\gamma_{-}$ and a drive with unit modulation amplitude, i.e. $\delta=1$. For the static case of this problem, the Liouvillian has a second-order EP at $\gamma_{-}=8J$, whereas the non-Hermitian Hamiltonian that results from postselecting on the quantum trajectories with no quantum jumps has a second-order EP at $\gamma_{-}=4J$. Figure~\ref{fig:jfloquet1}a shows the resultant $\mathrm{IP}(\gamma_{-},\Omega)$ map for a sinusoidal drive, Eq.(\ref{eq:jf}). When $\mathrm{IP}=1$, one pair among the 3 transient eigenmatrices coalesces and indicates an EP in the Lindblad dynamics. When $\Omega=0$, the static case, the Liouvillian EP at $\gamma_-=8J$ is observed~\cite{Minganti2019}. At the other end, when $\Omega/J\gg 1$, a line of EPs emerges at $\gamma_{-}=4J$ because the sinusoidal modulation averages to zero. Remarkably, we find that the EP contours extend down to zero dissipator strength ($\gamma_{-}=0$) at modulation frequencies $\Omega_n=2J/n$ for $n\geq 1$. Figure~\ref{fig:jfloquet1}b shows that for a square-wave modulation of the drive $J(t)$, the phase diagram is qualitatively similar, including the static-limit EP at $\gamma_{-}=8J$ and the high-frequency EP line at half the dissipator strength. Figure~\ref{fig:jfloquet1}c-f show high-resolution maps for a sinusoidal drive in the vicinity of even and odd subharmonics. In the neighborhood of $\Omega_n=2J/n$, the EPs contours are given by straight lines with slope $\pm 4n$, shown by white lines in Fig.~\ref{fig:jfloquet1}c-e. We also see that as the dissipator strength increases, the EP contours asymmetrically deviate from the straight-line prediction that is valid in the limit $\gamma_{-}/J\ll 1$. This $n$-dependent shift can be understood as the non-Hermitian version of the Bloch-Siegert shift~\cite{Lee2015} for multi-photon resonances in the traditional Rabi problem~\cite{Bloch1940,CohenT1973,CohenT1973a}. Figure~\ref{fig:jfloquet1}g shows representative, parametric evolution of the four eigenvalues $\lambda_k$ of $G(T)$ at $\gamma_{-}/J=0.4$ for three points, marked by triangles, in Fig.~\ref{fig:jfloquet1}e. Since the eigenvalues $\lambda_k$ are real or complex conjugates, and there is a trivial, steady-state eigenvalue, $\lambda_0=1$, the remaining three eigenvalues can either be all on the real axis ($\Omega/J=2$) or two are in the complex plane ($\Omega/J=1.9$ and $\Omega/J=2.17$). We call the region where all $\lambda_k$ are real as the overdamped region, whereas the region with complex-conjugate $\lambda_k$ is called the underdamped region. The contours of exceptional EPs (approximated by straight lines with slopes $\pm4$ in Fig.~\ref{fig:jfloquet1}e) separate these regions, and signal the fastest approach to the equilibrium state that occurs at the EP. It is also worth noting that the phase diagram remains the same when the dissipator $\sigma_{-}$ is changed to $\sigma_{+}$.

%------------------------------------------------%
%------------------------------------------------%
\begin{figure}[h]
\centering
\includegraphics[width=\columnwidth]{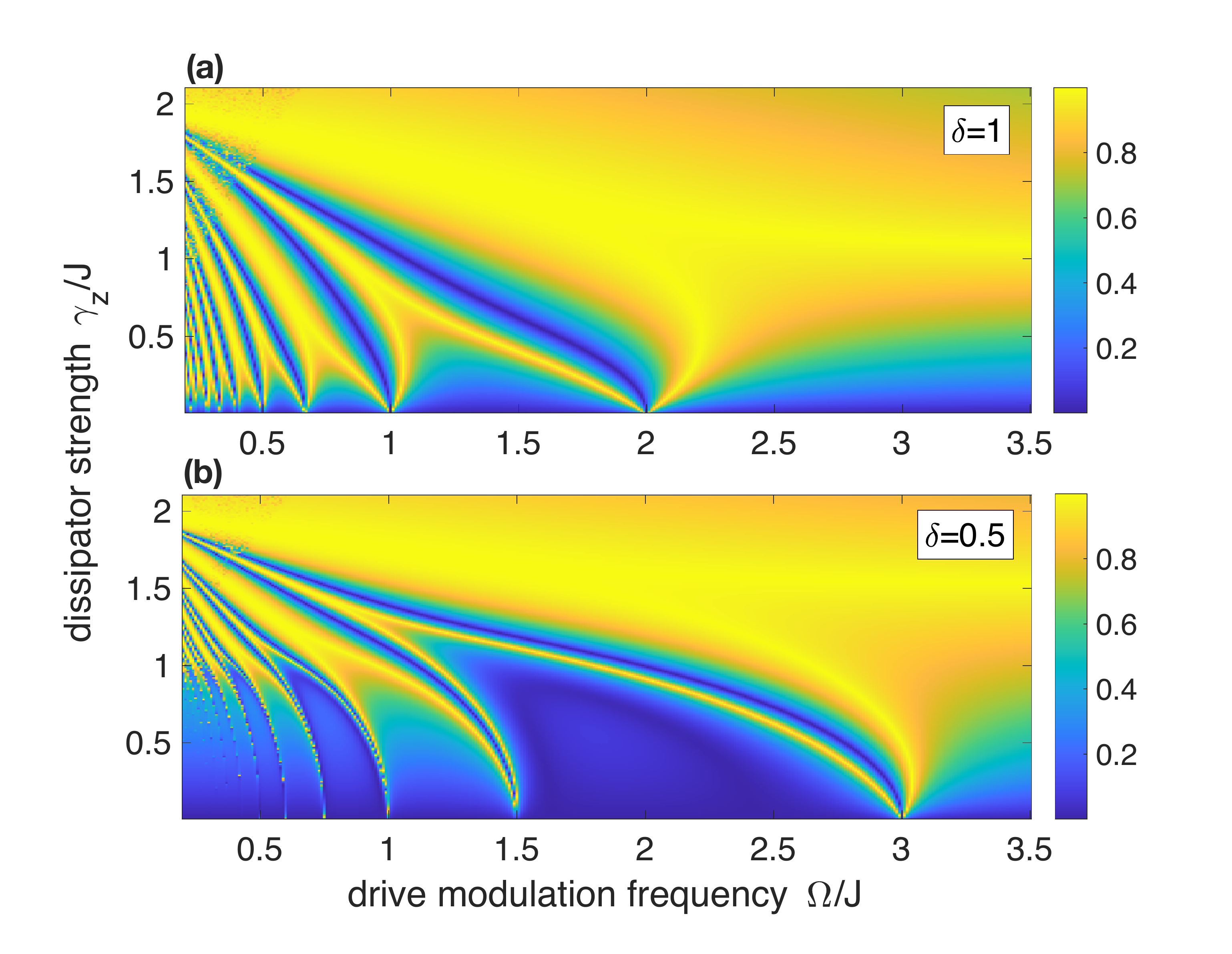}
\caption{EP contours for a qubit with sinusoidally modulated drive and phase noise dissipator. (a) At $\delta=1$, the $\mathrm{IP}(\gamma_z,\Omega)$ map shows the static EP at $\gamma_z=2J$, and an EP line at $\gamma_z=J$ in the high-frequency regime $\Omega/\gamma\gg 1$. Additionally, lines of EP merge on the $\Omega$-axis at $\Omega_n=2J/n$ for $n\geq 1$, giving rise to transitions between underdamped and overdamped transients in the Lindblad dyamics. (b) For a smaller modulation strength, $\delta=0.5$, the high-frequency EP line shifts to $\gamma_z=(2-\delta)J$, and the primary resonance shifts from $\Omega_1=2J$ at $\delta=1$ to $\Omega_1=2J(2-\delta)=3J$.}
\label{fig:jfloquet2}
\end{figure}
%-------------------------------------------------%
%-------------------------------------------------%

For the phase noise dissipator $F=\sigma_z$ with strength $\gamma_z$, the static Liouvillian has a second-order EP at $\gamma_z=2J$, whereas the postselected non-Hermitian Hamiltonian has no EPs. Figure~\ref{fig:jfloquet2} shows the EP phase diagram for the time-periodic case. For a sinusoidal drive with unit amplitude, Fig.~\ref{fig:jfloquet2}a shows that the inner-product map $\mathrm{IP}(\gamma_z,\Omega)$ has a static EP at the expected location, as well as a line of EPs at high frequency $\Omega/J\gg1$ at $\gamma_z=J$. In addition, we see the EP contours emerging from resonances at $\Omega_n=2J/n$ at vanishingly small dissipator strengths. The results, qualitatively, are similar to those in Fig.~\ref{fig:jfloquet1}. When the drive $J(t)$ is modulated with a smaller amplitude, $\delta=0.5$, the resultant $\mathrm{IP}$ map is shown in Fig.~\ref{fig:jfloquet2}b. The key differences are that the high-frequency EP contour shifts to the time-averaged $J(t)$ value of $\gamma_z=J(2-\delta)=1.5J$ and the fundamental resonance frequency for the EP on the $\Omega^+$-axis shifts from $\Omega_1=2J$ to $\Omega_1=2J(2-\delta)=3J$. Note that as $\delta\rightarrow 0$, the amplitude of the modulation goes to zero and, the overdamped regions become narrower, and as is expected, the Floquet results map onto to the static-limit results. On the other hand, when $\delta\rightarrow 2$, i.e. the drive modulation $J(t)$ averages to zero, $\Omega_1\rightarrow 0$ and the qubit is in the overdamped state for any dissipator strength. 
%----------------------------------------------------------------------------------------------------------------------%

\section{EPs with a periodic dissipation}
\label{sec:gfloquet}
In this section, we investigate the effect of periodically varying the dissipator strength $\gamma(t)$ in the presence of a constant drive $H_s=-J\sigma_x$. Here, $\gamma(t)$ varies either sinusoidally or in a square-wave format with period $T=2\pi/\Omega$, i.e 
\begin{equation}
\label{eq:gf}
\gamma(t)=\left\{\begin{array}{c}
(\gamma/2)[1+\cos(\Omega t)],\\
\gamma\,\mathrm{Square}(\Omega t).
\end{array}\right.
\end{equation}
It is worth its while to point out that the ``completely postitive'' aspect of the Lindblad dynamics is contingent on $\gamma(t)\geq 0$, in contrast to Eq.(\ref{eq:jf}) where the drive can undergo a $\pi$-phase change. We start with the spontaneous emission dissipator $F=\sigma_{-}$. In the static case, when the quantum jumps are ignored, the resulting non-Hermitian Hamiltonian has an EP at $\gamma_c=4J$.  Since the static ($\Omega/J\ll 1$) and high-frequency ($\Omega/J\gg 1$) limits at moderate-to-high dissipation strengths, i.e. $\gamma_{-}\gtrsim \gamma_c$, are clear, we will restrict ourselves to small dissipation strengths. 

\begin{figure}
\centering
\includegraphics[width=\columnwidth]{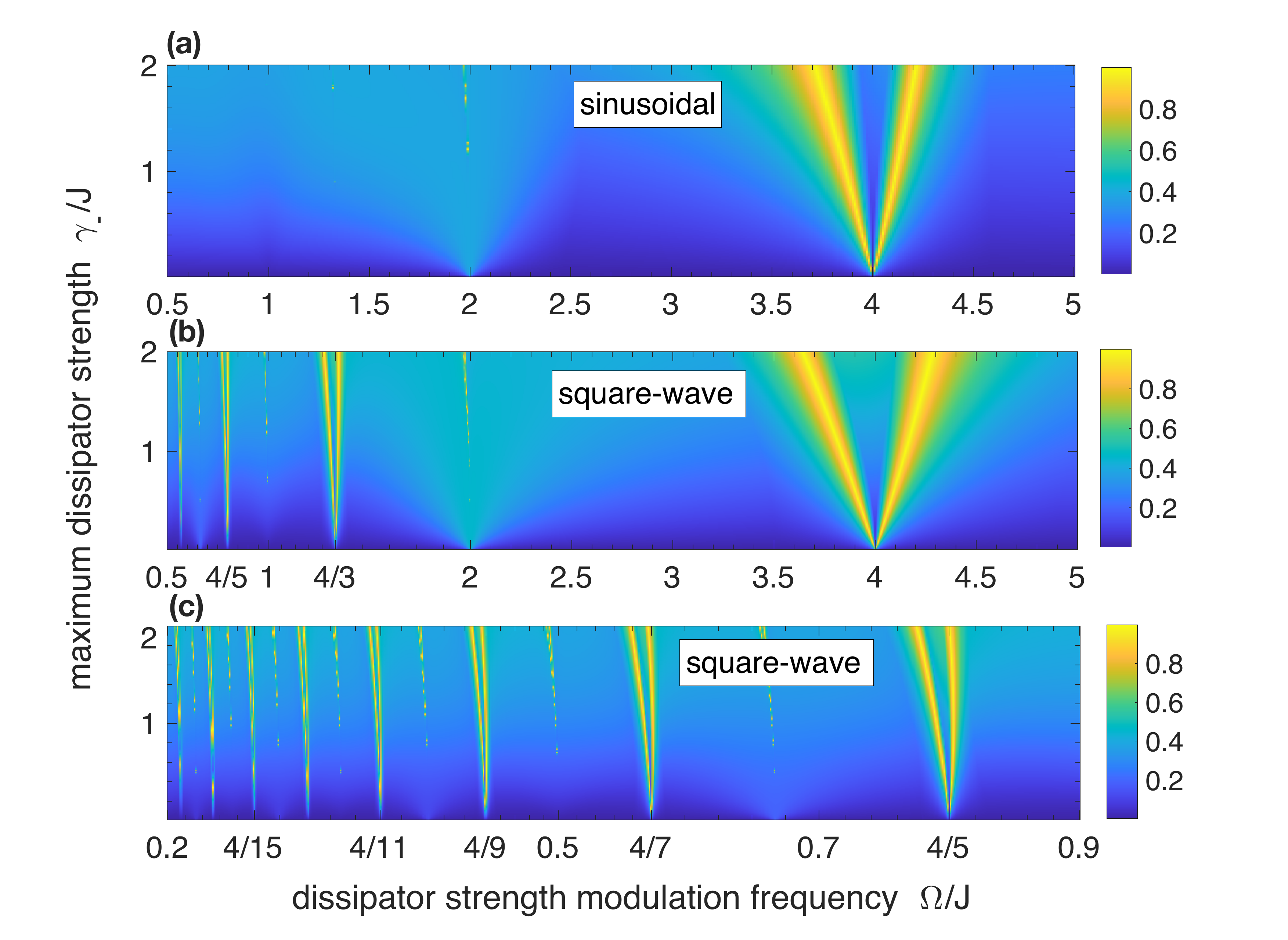}
\caption{$\mathrm{IP}(\gamma_{-},\Omega)$ map for $F=\sigma_{-}$. (a) For a sinusoidally modulated $\gamma(t)$, the $\mathrm{IP}$ map shows two lines of EPs converging at $\Omega_0=4J$ at vanishingly small $\gamma_{-}$, while EP contours at lower $\Omega_m=4J/(2m+1)$ are too thin to be seen. (b) For a square-wave modulation, additional EP contours at seen at higher subharmonics, as well as regions of enhanced $\mathrm{IP}$ at even subharmonics $\Omega'_m=4J/2m$. (c) Higher-resolution scans for square-wave modulation show that the width of the overdamped region between two EP lines converging at $\Omega_m$ scales linearly with $\gamma_{-}$, making their observation feasible. For a sinusoidal modulation, this width scales as $(\gamma_{-}/J)^{2m+1}$ and is thus much smaller than the present case~\cite{Lee2015,Li2019}.}
\label{fig:gfloquet1}
\vspace{-5mm}
\end{figure}

Figure~\ref{fig:gfloquet1} shows the resultant $\mathrm{IP}(\gamma_{-},\Omega)$ map. For a sinusoidal variation of $\gamma_{-}(t)$, two lines of EPs converge at vanishingly small $\gamma_{-}$ to $\Omega_0=4J$ (Fig.~\ref{fig:gfloquet1}a). A careful numerical analysis shows that there are odd, subharmonic resonances at $\Omega_m=4J/(2m+1)$ for $m\geq 1$, but those slivers are too thin to see on the scale shown here. In addition, we see enhanced non-orthogonality of the eigenmatrices along the $\Omega$-axis at even subharmonics $\Omega'_m=4J/2m$ with $m\geq 1$ although none of them reach an $\mathrm{IP}=1$ indicative of an EP. Figure~\ref{fig:gfloquet1}b shows corresponding results for a square-wave modulation. Here, the EP contours converging onto odd subharmonics $\Omega_0$, $\Omega_1=4J/3$ and $\Omega_2=4J/5$ are clear, as are triangular regions of enhanced $\mathrm{IP}$ value are seen at even subharmonics. A higher-resolution scan in Fig.~\ref{fig:gfloquet1}c shows these EP contours at eight lower subharmonics $\Omega_m$ ($m=5-13$). These results suggest that traversing the EP contours by changing modulation frequency at small dissipation strengths is more feasible with a square-wave modulation instead of the sinusoidal one, particularly when it comes to EP lines at higher subharmonics. They are consistent with results for the corresponding non-Hermitian Hamiltonian with modulated, mode-selective loss, where the width of the broken region scales as $(\gamma_{-}/J)^{2m+1}$ for sinusoidal modulation, whereas it scales linearly with $\gamma_{-}$ for a square-wave modulation~\cite{Lee2015,Li2019}. 

%------------------------------------------------%
%------------------------------------------------%
%-------------------------------------------------%
%-------------------------------------------------%

\begin{figure}[h]
\centering
\includegraphics[width=\columnwidth]{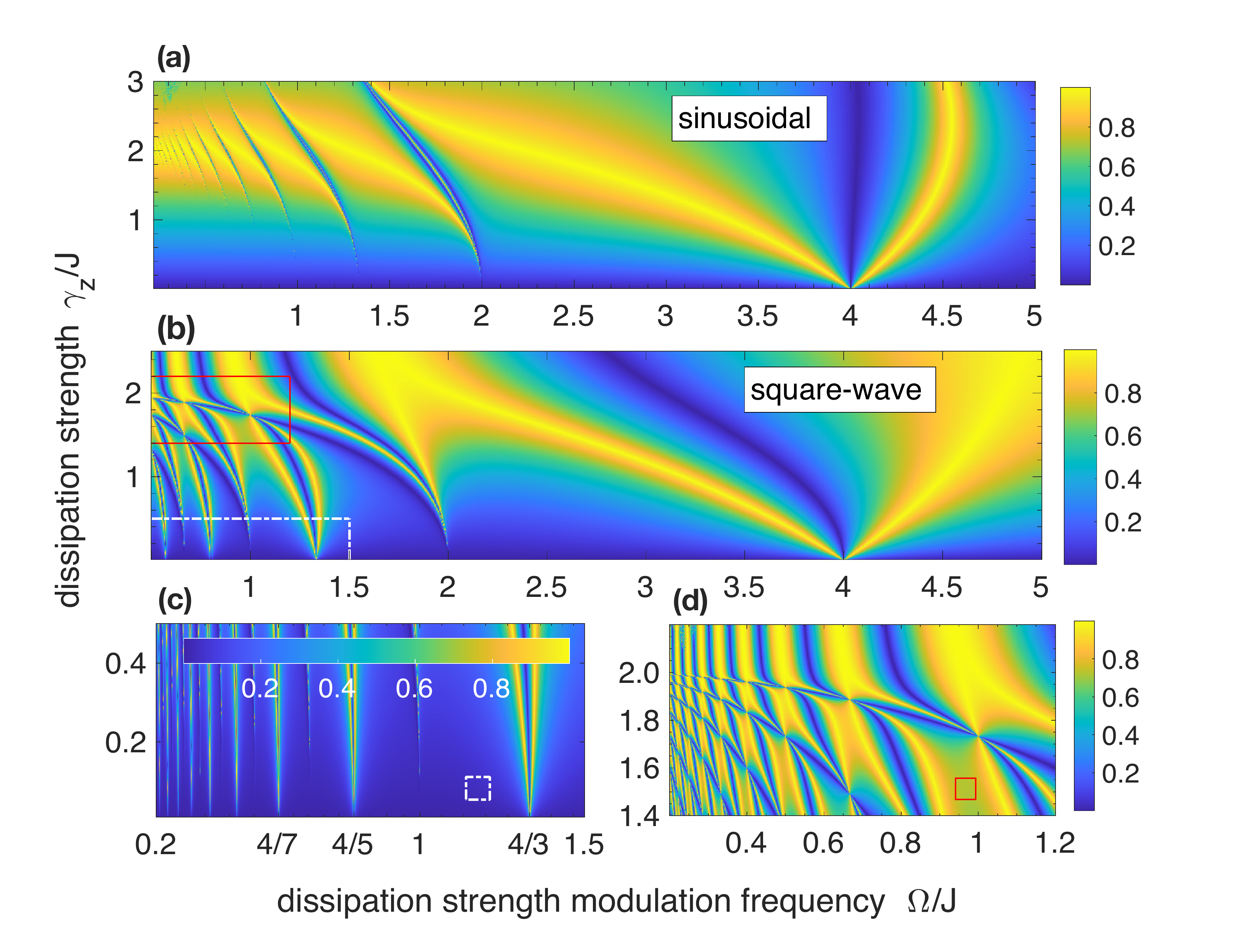}
\caption{$\mathrm{IP}(\gamma_z,\Omega)$ map for a system with phase noise, i.e. $F=\sigma_z$. (a) For sinusoidal modulation, EP contours converge at $\Omega_0=4J$, but higher subharmonics are not visible. Deep in the overdamped triangular region anchored at $\Omega_0$, the eigenmatrices are orthogonal resulting in $\mathrm{IP}=0$. (b) For a square-wave modulation, EP contours merging at odd subharmonics $\Omega_m=4J/(2m+1)$ are seen. (c) High-resolution scan shows that the width of overdamped region scales linearly with $\gamma_z$; 8 subharmonics can be seen. (d) High-resolution scan in the vicinity of the static EP, $\gamma_z=2J$, shows a rich structure with intersecting contours of EPs ($\mathrm{IP}=1$) and orthogonal eigenmatrices ($\mathrm{IP}=0$).}
\label{fig:gfloquet3}
\vspace{-3mm}
\end{figure}

Lastly, we consider $F=\sigma_z$, where the phase-noise strength $\gamma_z(t)$ is periodically varied. Figure~\ref{fig:gfloquet3} summarizes the results for this scenario. For a sinusoidally varying $\gamma_z(t)$, we see the static EP signature at $\gamma_{z}=2J$ at low frequencies.  At high frequencies, due to the temporal averaging, a line of EPs emerges at $\gamma_{z}=4J$. In addition, we see clear EP contour converging to $\Omega_0=4J$, while their presence at lower subharmonics is not unclear. Also remarkable is the nearly vertical line, emanating from $\Omega_0$ indicating $\mathrm{IP}\approx 0$. It shows that deep in the overdamped region, the eigenmatrices of Liouvillian become orthogonal, just as they are in the Hermitian ($\gamma_z=0$) limit. Figure~\ref{fig:gfloquet3}b shows the $\mathrm{IP}$ heat-map for a square-wave modulation. Here, we clearly see EP lines converging on odd subharmonics $\Omega_m$. A high-resolution scan at low frequencies and small $\gamma_z$ shows that the width of the overdamped region scales linearly with $\gamma_z$ (Fig.~\ref{fig:gfloquet3}c). 

Figure~\ref{fig:gfloquet3}d shows the high-resolution scan in the vicinity of the static EP. In addition to EP contours, it clearly shows contours with $\mathrm{IP}\approx 0$ that interweave with the EP contours. This rich Floquet structure in the vicinity of the static EP is a key feature of non-Hermitian Floquet problem. Here, in particular, it implies that with minor changes to the dissipator strength or modulation frequency, one can rapidly switch the trajectory of the system's approach to equilibrium between underdamped and overdamped. 
%------------------------------------------------%
%------------------------------------------------%
%-------------------------------------------------%
%-------------------------------------------------%

\section{Analytical Results}
\label{sec:exact}
In this section, we briefly present analytical results that encode some features of the numerically obtained EP phase diagrams. Since the Liouvillians $\mathcal{L}(t)$ at different instances of time do not, generally, commute with each other, it is not possible to obtain analytical expressions for the $4\times 4$ time-evolution operator matrix $G(T)$. In cases of a piecewise constant Hamiltonian or dissipator strengths, such as the square-wave modulation, $G(T)$ simplifies a bit, but still not enough for analytical calculations. Therefore, instead of the vectorized density-matrix representation of the system, for the single qubit case, we use the Bloch representation where $\rho(t)=(\mathbbm{1}_2+{\bf s}\cdot{\bf \sigma})/2$, ${\bf \sigma}=(\sigma_x,\sigma_y,\sigma_z)$ is the standard Pauli matrix vector, and $s_\alpha=\Tr(\rho\sigma_\alpha)$ denotes the $\alpha$-th component of the density-matrix spin projection. The Lindblad dynamics of the dissipative qubit can be cast as an inhomogeneous Bloch equation~\cite{AmShallem2015} $d{\bf s}/dt=A(t){\bf s}+b(t)$ where $b(t)=[0,0,\gamma_{-}(t)]^T$ and
\begin{equation}
\label{eq;blocha}
A(t)=-\left[\begin{array}{ccc}
\frac{\gamma_{-}(t)}{2}+2\gamma_z(t) & 0 & 0 \\
0 & \frac{\gamma_{-}(t)}{2}+2\gamma_z(t) & -2J(t)\\
0 & 2J(t) & \gamma_{-}(t)
\end{array}\right].
\end{equation}
Since $A$ is block-diagonal, so is the Bloch time-evolution operator
\begin{equation}
\label{eq:blochb}
B(T)=\mathbb{T} e^{\int_0^T A(t') dt'}=\left[\begin{array}{cc} \lambda_1 & 0 \\ 0 & B_2(T)\end{array}\right],
\end{equation}
which encodes the information about long-time dynamics of the system. It is obvious that $\lambda_1=\exp\left[\int_0^T dt' A_{11}(t')\right]<1$ is one eigenvalue of $B(T)$ (or equivalently, $G(T)$). The remaining two eigenvalues $\lambda_{2,3}$ are determined by the $2\times 2$ matrix $B_2(T)$. For a square-wave modulation of the drive $J(t)$ and $F=\sigma_{-}$, the EP contours, defined by $\lambda_2=\lambda_3$, are determined by 
\begin{align}
\label{eq:b2}
\sinh\left[\frac{\pi\gamma_{-}}{4\Omega}\right]\cos\alpha\cos\beta+\cosh\left[\frac{\pi\gamma_{-}}{4\Omega}\right]\sin\alpha\sin\beta=\pm\sin\beta,
\end{align}
where $\gamma_{-}=8J\sin\alpha$ and $\beta=2\pi J\cos\alpha/\Omega$. In the limit $\alpha\rightarrow 0$, Eq.(\ref{eq:b2}) is satisfied with $\Omega=\Omega_n=2J/n$ for $n\geq 1$. On the other hand, when $\Omega\rightarrow0$, the static EP at $\gamma_{-}=8J$ is recovered. Perturbative expansion around $\Omega=\Omega_n$ gives that the two EP lines emerging from that point have slope $\pm 4n$, as shown in Fig.~\ref{fig:jfloquet1}c-e. When the static dissipator is phase noise, a square-wave drive leads to EPs at $\Omega_m=2J/m$, with the slopes of the two emerging EP lines given by $\pm m$ (Fig.~\ref{fig:jfloquet2}). 

A similar analysis for square-wave modulation of dissipator strengths, with $\sigma_{-}$ and $\sigma_z$ as dissipators, gives lines of EPs converging onto the $\Omega$-axis at odd subharmonic resonance values $\Omega_m=4J/(2m+1)$ for $m\geq 0$. For the $\sigma_{-}$ dissipator, a perturbative analysis shows that the slopes of EP lines emerging from $\Omega_m$ is given by $\pm2\pi(2m+1)^2$ (Fig.~\ref{fig:gfloquet1}). For the phase-noise dissipator $\sigma_z$, the corresponding slopes are reduced to a quarter, i.e. they are given by $\pm\pi (2m+1)^2/2$ (Fig.~\ref{fig:gfloquet3}). 

%----------------------------------------------------------------------------------------------------------------------%

\section{Discussion} 
\label{sec:disc}
In this paper, we have investigated the exceptional points of Lindblad dynamics with time periodic parameters. By mapping an $N$-dimensional superoperator problem onto an $N^2$-dimensional Liouvillian, we have analyzed the eigenvalues, and non-orthogonality of corresponding eigenmatrices for the time evolution operator $G(T)$ for the Liovillian. For the static case and in the high-frequency limit, the EPs occur when the time-averaged drive strength is comparable to the time-averaged dissipator strength. Surprisingly, however, we also find that EPs occur at vanishingly small dissipator strengths when the modulation frequency matches the primary resonance or a suitable subharmonic of the relevant energy scale $\sim J$. These results are robust, in that they do not depend upon the exact form of the temporal variation, only on its periodic nature. 

In this work, we have only considered the EP contours of a single qubit. Although this might seem limiting, in reality, it is not. The occurrence of a second-order EP corresponds to the coalescence of two eigenmatrices of the Liouvillian matrix. For higher-dimensional system such as a qutrit ($N=3$) or two interacting qubits ($N=4$), the Liouvillian is an $N^2$-dimensional matrix. However, its EP structure is still determined by the coalescence of two eigenmatrices. Therefore, it is sufficient to consider the minimal model, just as it is sufficient to confine oneself to two relevant levels when we study second-order EPs of effective, non-Hermitian Hamiltonians.

Finally, let us consider the consequences of a periodic drive or dissipation on time evolution. In the static case~\cite{Minganti2019}, the transition across the EP can be observed in the trajectory of ${\bf s}(t)$ to the equilibrium state: it changes from underdamped to overdamped. In contrast, for the time-periodic case, even when the eigenvalues $\lambda_k$ of the time-evolution operator $G(T)$ are purely real, the non-stroboscopic  spin-projection dynamics ${\bf s}(t)$ (at times $t\neq nT$) has oscillations in both ``underdamped'' and ``overdamped'' regions. Such oscillations can only be removed by sampling the density-matrix dynamics of a weakly dissipative qubit ($\gamma/J\sim 10^{-3}$) stroboscopically for times $t\gg T$. These considerations suggest that an experimental investigation of transiting across the Floquet EP contours by changing the time-modulation frequency is feasible, but challenging, in the weak dissipation limit. 
%----------------------------------------------------------------------------------------------------------------------%

\bibliographystyle{apsrev4-1}
\bibliography{pt_bib.bib}
\end{document}